\documentclass[twocolumn,aps,prb,10pt,superscriptaddress,longbibliography]{revtex4-2}

\usepackage{amsmath}
\usepackage{amssymb}
\usepackage{graphicx}
\usepackage{dcolumn}
\usepackage{bm}
\usepackage{xcolor}
\usepackage[colorlinks,linkcolor=blue,citecolor=blue,urlcolor=blue]{hyperref}
\usepackage[markup=underlined]{changes}

\begin{document}

\title{Knudsen-Controlled Switching of Thermal Conductivity Response by Targeted Phonon Excitation}

\author{Shixian Liu}
\affiliation{Department of Thermophysics, Bauman Moscow State Technical University, Moscow, 105005, Russia}

\author{Tianhao Li}
\affiliation{College of Science, National University of Defense Technology, Changsha 410073, China}
\affiliation{School of Energy and Power Engineering, Huazhong University of Science and Technology, Wuhan, 430074, China}

\author{Fei Yin}
\affiliation{Department of Thermophysics, Bauman Moscow State Technical University, Moscow, 105005, Russia}

\author{Yu He}
\affiliation{College of Science, National University of Defense Technology, Changsha 410073, China}

\author{Alexander A. Barinov}
\affiliation{Department of Thermophysics, Bauman Moscow State Technical University, Moscow, 105005, Russia}

\author{Han Meng}
\affiliation{College of Science, National University of Defense Technology, Changsha 410073, China}

\author{Nuo Yang}
\email{nuo@nudt.edu.cn}
\affiliation{College of Science, National University of Defense Technology, Changsha 410073, China}

\date{\today}

\begin{abstract}
Targeted phonon excitation offers a route to dynamically control heat conduction, yet no general principle predicts whether a spectrally selective nonequilibrium phonon population will enhance or suppress thermal transport. A Knudsen-controlled competition between the increased contribution of long-mean-free-path phonons and excitation-enhanced intrinsic scattering governs the sign of the thermal-conductivity response. First-principles three-phonon scattering rates combined with phonon-tracking Monte Carlo simulations are used to examine Ge, Si, and 3C--SiC from bulk crystals to confined nanofilms. In bulk systems, excitation-enhanced scattering dominates and thermal conductivity is predominantly suppressed. In nanofilms, by contrast, low-frequency excitation can increase the contribution of quasi-ballistic heat-carrying channels and enhance thermal conductivity, whereas higher-frequency excitation is predominantly suppressive. At fixed background temperature and excitation strength, these opposite responses are organized in a frequency--Knudsen map based on the normalized target frequency, $\omega_{\mathrm t}/\omega_{\mathrm D}$, and the Knudsen number, $\mathrm{Kn}$. The resulting framework provides a general physical basis for controlling nonequilibrium heat transport beyond static phonon engineering.
\end{abstract}

\maketitle


Controlling heat conduction in solids is a longstanding goal in condensed-matter physics and materials engineering. The ability to manipulate phonon transport is increasingly important for thermal management in high-performance and artificial-intelligence electronics, optoelectronic and photonic devices, advanced energy-conversion systems, and biointegrated sensing technologies~\cite{woon_thermal_2025,qian_phonon-engineered_2021,maldovan_sound_2013,jiang_inorganic_2018}.

Conventional phonon engineering addresses this goal through nanostructuring~\cite{qian_phonon-engineered_2021,nomura_thermal_2018,ma_unexpected_2018,liu_quantifying_2025}, defect engineering~\cite{lee_mechanism_2012,wei_influence_2023,wu_isotope_2024}, and strain modulation~\cite{xie_first-principles_2025,qin_significant_2025,zhang_strain-gradient-driven_2026}. These approaches modify phonon spectra, lifetimes, or scattering phase space and have proved effective in tailoring, and often suppressing, thermal conductivity. However, because the resulting transport properties are largely encoded in the material structure, heat conduction remains essentially passive once the system has been fabricated. Driving phonons out of equilibrium offers a fundamentally different route in which thermal transport can be controlled through the phonon population itself~\cite{li_colloquium_2012,wehmeyer_thermal_2017,aryana_observation_2022,liu_actively_2024,dong2022phonon}.

Recent experiments have demonstrated that nonequilibrium phonon populations can be generated and manipulated using resonant mid-infrared and terahertz fields. These include mode-selective excitation of an optical phonon in SiC, terahertz control of phonon-mediated carrier relaxation, and anharmonic energy transfer between phonon modes~\cite{yoshida_experimental_2013,sekiguchi_enhancing_2021,kozina_terahertz-driven_2019}. Non-equilibrium phonon polaritons have also been shown to produce measurable enhancement of heat conduction in SiC nanowires~\cite{pan_remarkable_2023}. Related advances in phonon-specific structural control and electrically generated phonon polaritons further broaden the experimental routes for creating driven phonon states~\cite{jia_optically_2026,abou-hamdan_electroluminescence_2025,guo_hyperbolic_2025}.

How spectrally selective phonon excitation controls heat conduction, however, remains poorly understood. Nonthermal phonon populations have been predicted to enhance heat transport at micro- and nanoscale dimensions~\cite{chiloyan_thermal_2020}. Recent studies of low-dimensional materials, including graphene~\cite{wan_modulating_2024} and hexagonal boron nitride~\cite{pan_using_2025}, have further shown that selectively exciting low-frequency phonons can increase, rather than suppress, thermal conductivity. These findings challenge the conventional expectation that an excess phonon population should primarily strengthen phonon--phonon scattering and reduce thermal conductivity. Instead, targeted excitation produces two competing effects: it increases the transport weight of selected phonon modes while simultaneously enhancing their intrinsic scattering.

The central unresolved question is therefore what determines whether targeted phonon excitation enhances or suppresses heat conduction. Exciting low-frequency phonons can increase the contribution of modes with large group velocities and long intrinsic mean free paths. In bulk-like systems, however, the accompanying increase in three-phonon scattering directly shortens these mean free paths and can overwhelm the additional transport weight. Geometric confinement changes this balance. When the characteristic system size becomes comparable to the phonon mean free paths, boundary limitation reduces the relative transport penalty associated with additional intrinsic scattering, allowing the enhanced contribution of long-mean-free-path modes to become dominant. The response should therefore depend jointly on the spectral position of the excited phonons and the degree of confinement, naturally characterized by the target frequency and the Knudsen number~\cite{yang_mean_2013,chen_non-fourier_2021,liu_determination_2024,beardo_nanoscale_2025}.

This work combines first-principles three-phonon scattering rates with phonon-tracking Monte Carlo simulations to investigate targeted excitation in Ge, Si, and 3C--SiC from bulk crystals to confined nanofilms. The calculations show that bulk systems are predominantly scattering dominated, so targeted excitation mainly suppresses thermal conductivity. In nanofilms, by contrast, low-frequency excitation can increase the contribution of quasi-ballistic heat-carrying modes and produce a positive thermal-conductivity response, whereas higher-frequency excitation is predominantly suppressive. Despite their different phonon spectra and intrinsic scattering strengths, the three materials exhibit a common response-map topology when expressed in terms of the normalized target frequency, $\omega_{\mathrm t}/\omega_{\mathrm D}$, and the Knudsen number, $\mathrm{Kn}$. A reduced spectral transport model (RSTM) clarifies the physical origin of the sign reversal by showing how geometric confinement modifies the balance between excitation-induced spectral activation and enhanced intrinsic phonon--phonon scattering.

\section*{Results}

\subsection*{Physical picture of targeted phonon excitation}

\begin{figure}[htbp]
    \centering
    \includegraphics[width=\linewidth]{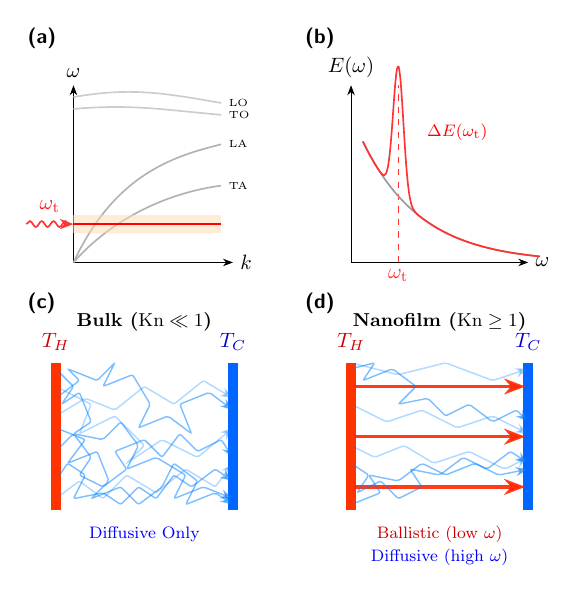}
    \caption{
    \textbf{Schematic illustration of targeted phonon excitation and the resulting transport regimes.}
    (a) Mode-selective excitation in the phonon dispersion, where energy is injected near the target angular frequency $\omega_{\mathrm t}$.
    (b) Nonequilibrium energy distribution $E(\omega)$ with an excess population near $\omega_{\mathrm t}$ relative to equilibrium.
    (c) Bulk-like transport regime, $\mathrm{Kn}\ll1$, in which excitation-enhanced intrinsic scattering dominates the response.
    (d) Confined transport regime, $\mathrm{Kn}\geq1$, in which the characteristic intrinsic phonon mean free path is comparable to or exceeds the film thickness. Boundary limitation reduces the relative influence of excitation-enhanced intrinsic scattering and allows low-frequency, long-mean-free-path phonons to retain a substantial quasi-ballistic contribution.
    }
    \label{fig1}
\end{figure}

The thermal-conductivity response to targeted phonon excitation arises from the competition between two excitation-induced effects, as illustrated in Fig.~\ref{fig1}. Energy is selectively injected into a narrow phonon-frequency window centered at the target angular frequency $\omega_{\mathrm t}$ [Fig.~\ref{fig1}(a)], producing a nonequilibrium population excess in the phonon energy distribution [Fig.~\ref{fig1}(b)]. This excess population influences heat transport through two opposing mechanisms. First, selectively populating low-frequency phonons with long intrinsic mean free paths increases their contribution to heat conduction. In confined geometries, these modes can retain a substantial quasi-ballistic contribution to the heat flux. Second, the same nonequilibrium population enhances intrinsic three-phonon scattering, thereby shortening phonon lifetimes and transport mean free paths. Whether thermal conductivity is enhanced or suppressed is therefore determined by the balance between these two effects.

The degree of geometric confinement is characterized by the Knudsen number,
\begin{equation}
\mathrm{Kn} = \frac{\langle\ell^0\rangle}{H},
\label{eq:Kn}
\end{equation}
where $H$ is the film thickness and $\langle\ell^0\rangle$ is the characteristic intrinsic phonon mean free path evaluated from the equilibrium phonon properties at the background lattice temperature. Here, $\langle\ell^0\rangle$ represents a system-level average and is not resolved with respect to phonon frequency.

In bulk crystals or sufficiently thick films, for which $\mathrm{Kn}\ll1$, heat transport is predominantly governed by intrinsic phonon--phonon scattering [Fig.~\ref{fig1}(c)]. Under these conditions, excitation-enhanced intrinsic scattering shortens the phonon transport mean free paths and therefore tends to suppress thermal conductivity. When $\mathrm{Kn}\geq1$, the characteristic intrinsic phonon mean free path is comparable to or larger than the film thickness, and boundary limitation becomes important [Fig.~\ref{fig1}(d)]. In this confined regime, selectively exciting low-frequency phonons can increase their quasi-ballistic contribution sufficiently to compensate for, or even overcome, the accompanying enhancement of intrinsic scattering.

\subsection*{Response in bulk crystals and nanofilms}

The thermal conductivities in the equilibrium reference state and under targeted phonon excitation are denoted by $\kappa_0$ and $\kappa$, respectively. The relative thermal-conductivity response is defined as
\begin{equation}
\frac{\kappa-\kappa_0}{\kappa_0}.
\label{eq:relative_response}
\end{equation}
Positive and negative values indicate excitation-induced enhancement and suppression, respectively.

\begin{figure*}[htb]
    \centering
    \includegraphics[width=0.85\linewidth]{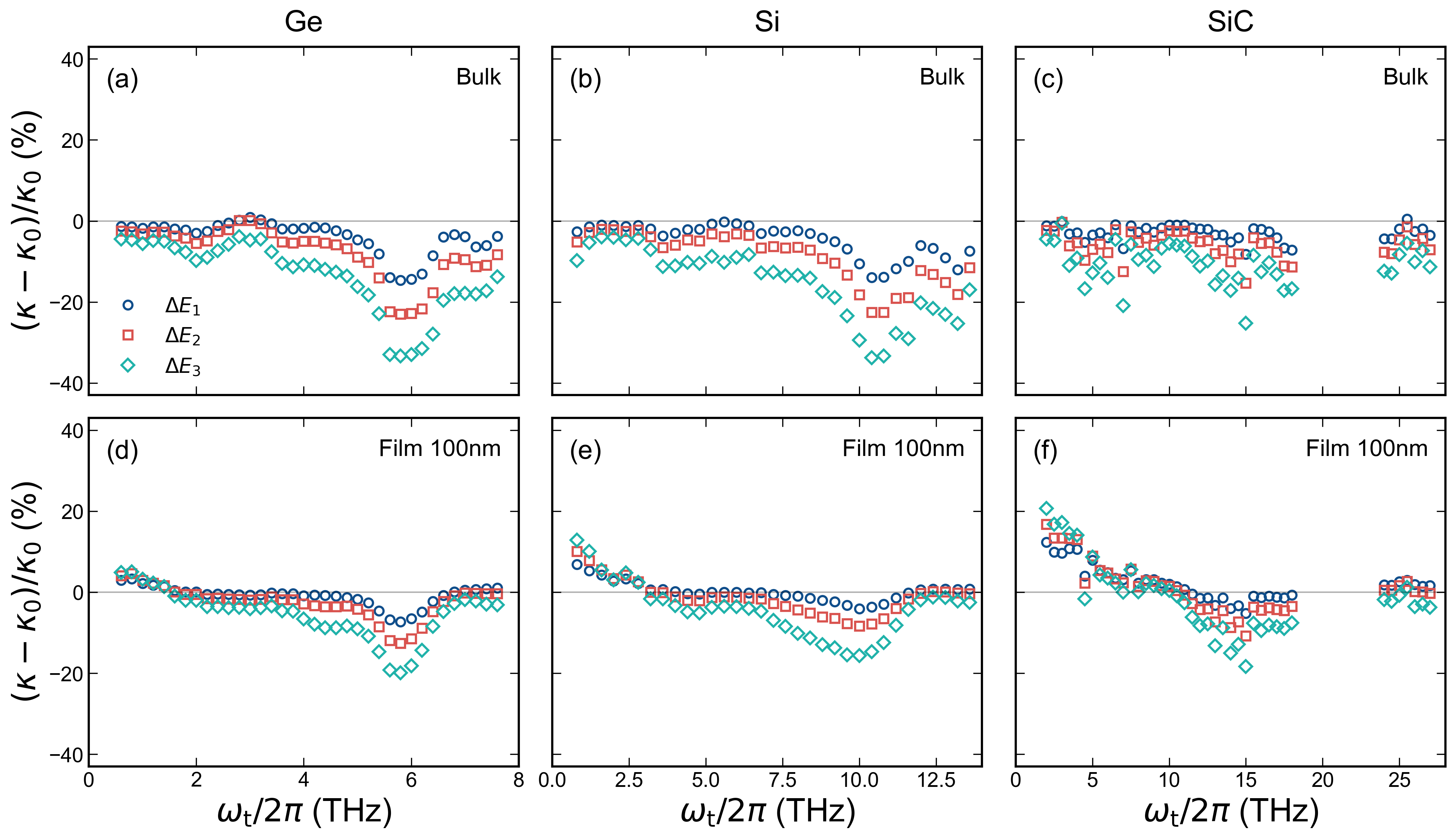}
    \caption{
    \textbf{Frequency-dependent switching of the thermal-conductivity response under targeted phonon excitation.}
    The relative thermal-conductivity change $(\kappa-\kappa_0)/\kappa_0$ is plotted as a function of the target frequency $\omega_{\mathrm t}/2\pi$ for bulk (a) Ge, (b) Si, and (c) 3C--SiC, and for the corresponding 100-nm films of (d) Ge, (e) Si, and (f) 3C--SiC.
    All results are evaluated at a background lattice temperature of $T=300~\mathrm{K}$. Circles, squares, and diamonds denote injected energy densities $\Delta E_1=5.0\times10^6~\mathrm{J\,m^{-3}}$, $\Delta E_2=1.0\times10^7~\mathrm{J\,m^{-3}}$, and $\Delta E_3=2.0\times10^7~\mathrm{J\,m^{-3}}$, respectively.
    Positive values indicate excitation-induced enhancement, whereas negative values indicate suppression. Bulk systems are predominantly scattering dominated and exhibit mainly negative responses. In contrast, 100-nm films show positive responses at low target frequencies and negative responses at higher frequencies, demonstrating confinement-induced switching between enhanced quasi-ballistic transport and scattering-dominated suppression.
    }
    \label{fig2}
\end{figure*}

The target frequency is reported as $\omega_{\mathrm t}/2\pi$ in units of THz. The frequency-dependent thermal-conductivity responses of bulk Ge, Si, and 3C--SiC and those of the corresponding 100-nm films, all evaluated at a background lattice temperature of $T=300~\mathrm{K}$, are compared in Fig.~\ref{fig2}.

For bulk Ge, Si, and 3C--SiC [Figs.~\ref{fig2}(a)--\ref{fig2}(c)], the response is predominantly negative over the investigated target-frequency range. This result shows that, in the bulk limit, targeted excitation mainly increases intrinsic phonon--phonon scattering rather than producing a sufficient increase in the effective heat-carrying contribution of the excited modes.

The suppression becomes stronger as the injected energy density increases from $\Delta E_1$ to $\Delta E_3$, demonstrating that the magnitude of the response is controlled by the strength of the nonequilibrium population excess. For Ge and Si, the largest reduction occurs in the intermediate-to-high-frequency range and reaches approximately $-30\%$ under the strongest excitation. For 3C--SiC, the suppressive response is distributed over a broader frequency range but likewise becomes stronger with increasing excitation energy.

A qualitatively different behavior appears in the corresponding 100-nm films [Figs.~\ref{fig2}(d)--\ref{fig2}(f)]. Low-frequency excitation produces a positive response in all three materials. Under the strongest excitation, the maximum enhancement is approximately $+5\%$ for Ge, $+10\%$ for Si, and $+20\%$ for 3C--SiC. This trend indicates that targeted excitation increases the transport weight of low-frequency, long-mean-free-path phonons whose contribution remains partly quasi-ballistic in the confined geometry.

As the target frequency increases, the enhancement weakens and eventually changes to suppression. At intermediate and high frequencies, the increase in intrinsic scattering outweighs the additional spectral heat-carrying weight, and the response becomes negative. The comparison between the upper and lower rows of Fig.~\ref{fig2} therefore demonstrates that geometric confinement can reverse the sign of the excitation-induced thermal-conductivity response.

\subsection*{Thickness and temperature dependence}

\begin{figure*}[htb]
    \centering
    \includegraphics[width=0.85\linewidth]{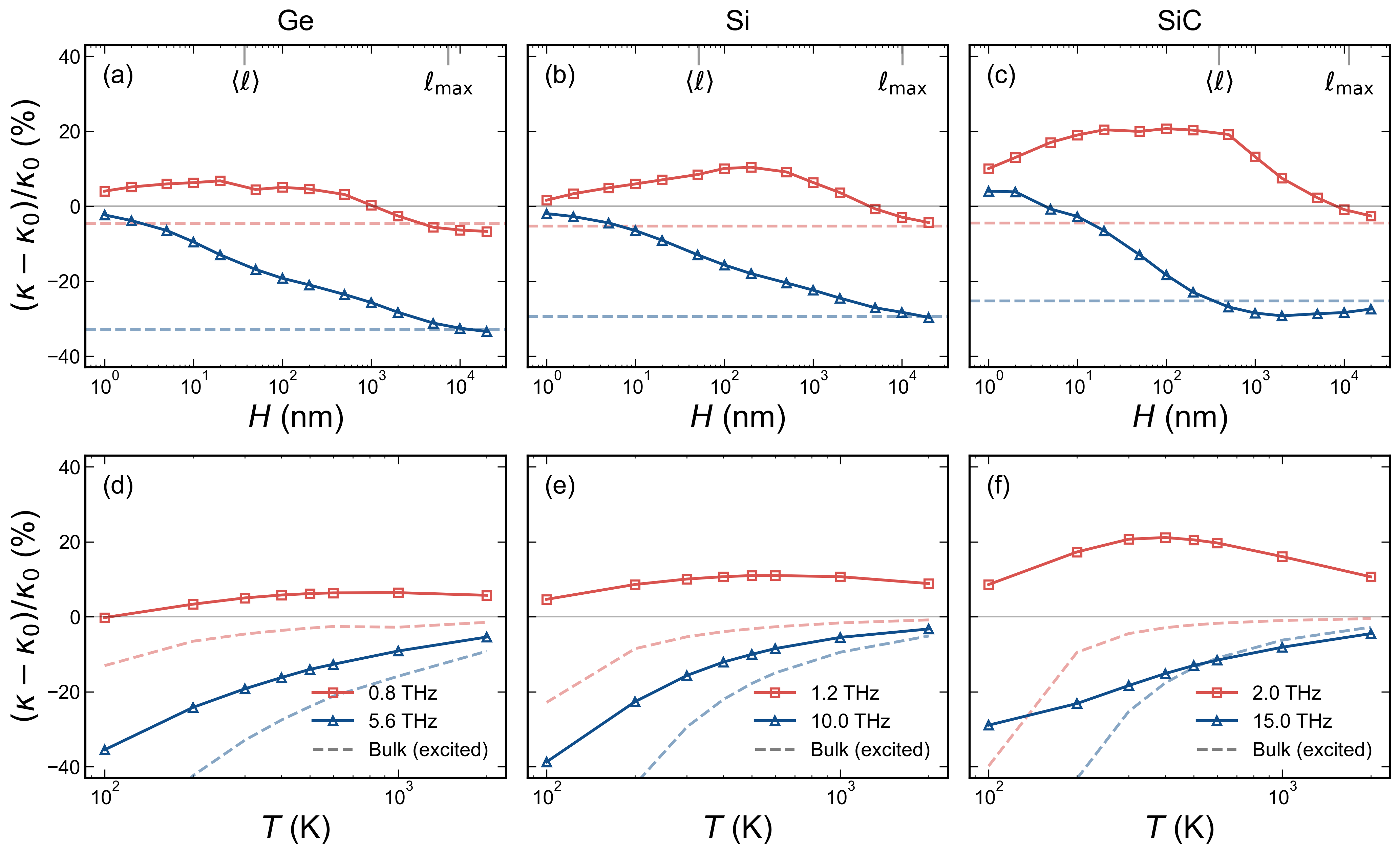}
    \caption{
    \textbf{Thickness and temperature dependence of the thermal-conductivity response.} All results are obtained at the injected energy density $\Delta E_3=2.0\times10^7~\mathrm{J\,m^{-3}}$.
    (a)--(c) Relative response $(\kappa-\kappa_0)/\kappa_0$ as a function of film thickness $H$ for representative low- and high-frequency excitations in Ge, Si, and 3C--SiC. The selected target frequencies $\omega_{\mathrm t}/2\pi$ are 0.8 and 5.6 THz for Ge, 1.2 and 10.0 THz for Si, and 2.0 and 15.0 THz for 3C--SiC, respectively.
    Horizontal dashed lines indicate the corresponding bulk-excited limits, and vertical dashed lines mark the average intrinsic mean free path $\langle\ell^0\rangle$ and maximum intrinsic mean free path $\ell_{\max}$.
    Low-frequency excitation exhibits a nonmonotonic thickness dependence, with the largest enhancement occurring when the film thickness is comparable to the characteristic phonon mean free path. High-frequency excitation is predominantly suppressive and approaches the bulk scattering-dominated limit as $H$ increases.
    (d)--(f) Temperature-dependent response in 100-nm films. Low-frequency excitation remains enhancing over a broad temperature range, whereas high-frequency excitation remains predominantly suppressive.
    }
    \label{fig3}
\end{figure*}

Figure~\ref{fig3} shows how film thickness and temperature control the competition between excitation-induced spectral activation and enhanced intrinsic scattering. All results are obtained at the injected energy density $\Delta E_3=2.0\times10^7~\mathrm{J\,m^{-3}}$.

Figures~\ref{fig3}(a)--\ref{fig3}(c) directly examine the role of geometric confinement by varying the film thickness. Under low-frequency excitation, all three materials exhibit a nonmonotonic response as $H$ increases. In the strongly confined regime, the response is positive or weakly positive because the excited low-frequency modes retain a quasi-ballistic contribution over the characteristic film dimension.

As $H$ increases, the enhancement initially grows and reaches a maximum when the film thickness becomes comparable to the characteristic intrinsic mean free path. The maximum enhancement is material dependent and reaches approximately $6\%$ in Ge, $10\%$ in Si, and $20\%$ in 3C--SiC. When the film thickness is increased further toward the bulk-like regime, the positive response decreases and eventually becomes negative, indicating that intrinsic scattering enhancement becomes dominant.

The enhancement is therefore not maximized in either limiting regime. In ultrathin films, equilibrium transport is already strongly boundary limited, and the available heat-carrying contribution of long-mean-free-path modes is substantially constrained. The additional gain produced by excitation is therefore modest. In thick films, boundary limitation becomes weak, and the excitation-induced reduction in intrinsic lifetimes directly suppresses transport. The largest enhancement appears in an intermediate Knudsen regime, where confinement weakens the relative penalty of additional intrinsic scattering without completely suppressing the contribution of long-mean-free-path phonons.

High-frequency excitation exhibits a different thickness dependence. For Ge and Si, the response remains negative over the investigated thickness range and approaches the corresponding bulk suppression limit as $H$ increases. The suppression grows from only a few percent in the strongly confined regime to approximately $30\%$ in the large-$H$ limit. High-frequency excitation provides little additional transport weight to long-propagating modes and instead predominantly enhances the intrinsic scattering experienced by the heat-carrying phonon population.

For 3C--SiC, a weak positive high-frequency response appears only under extreme confinement, after which the response rapidly changes to suppression as the film thickness increases. Thus, except in the most strongly confined regime, high-frequency excitation remains predominantly scattering dominated.

Figures~\ref{fig3}(d)--\ref{fig3}(f) show the temperature dependence of the response in 100-nm films. Under low-frequency excitation, the response remains positive over a broad temperature range in all three materials. The enhancement generally increases from low temperature to an intermediate temperature range and then decreases at higher temperatures.

At low temperature, the intrinsic mean free paths of the relevant acoustic phonons substantially exceed the film thickness, placing the 100-nm films in a strongly boundary-limited regime. Their effective propagation lengths are therefore already truncated by the geometry, which limits the additional transport gain produced by excitation. As temperature increases, the intrinsic mean free paths approach the film thickness, producing an intermediate-confinement regime in which the increased transport weight of the excited modes is most effective. At still higher temperatures, stronger intrinsic phonon--phonon scattering weakens the enhancement.

In contrast, high-frequency excitation remains suppressive over most of the temperature range. At low temperature, the imposed excess population constitutes a comparatively large perturbation to the weakly occupied high-frequency modes and strongly modifies the occupation factors entering the three-phonon scattering probabilities. As temperature increases, the same injected energy produces a smaller relative perturbation to the thermally populated phonon system, and the suppressive response weakens.

\subsection*{Knudsen--frequency phase diagram}

\begin{figure*}[htb]
    \centering
    \includegraphics[width=0.75\linewidth]{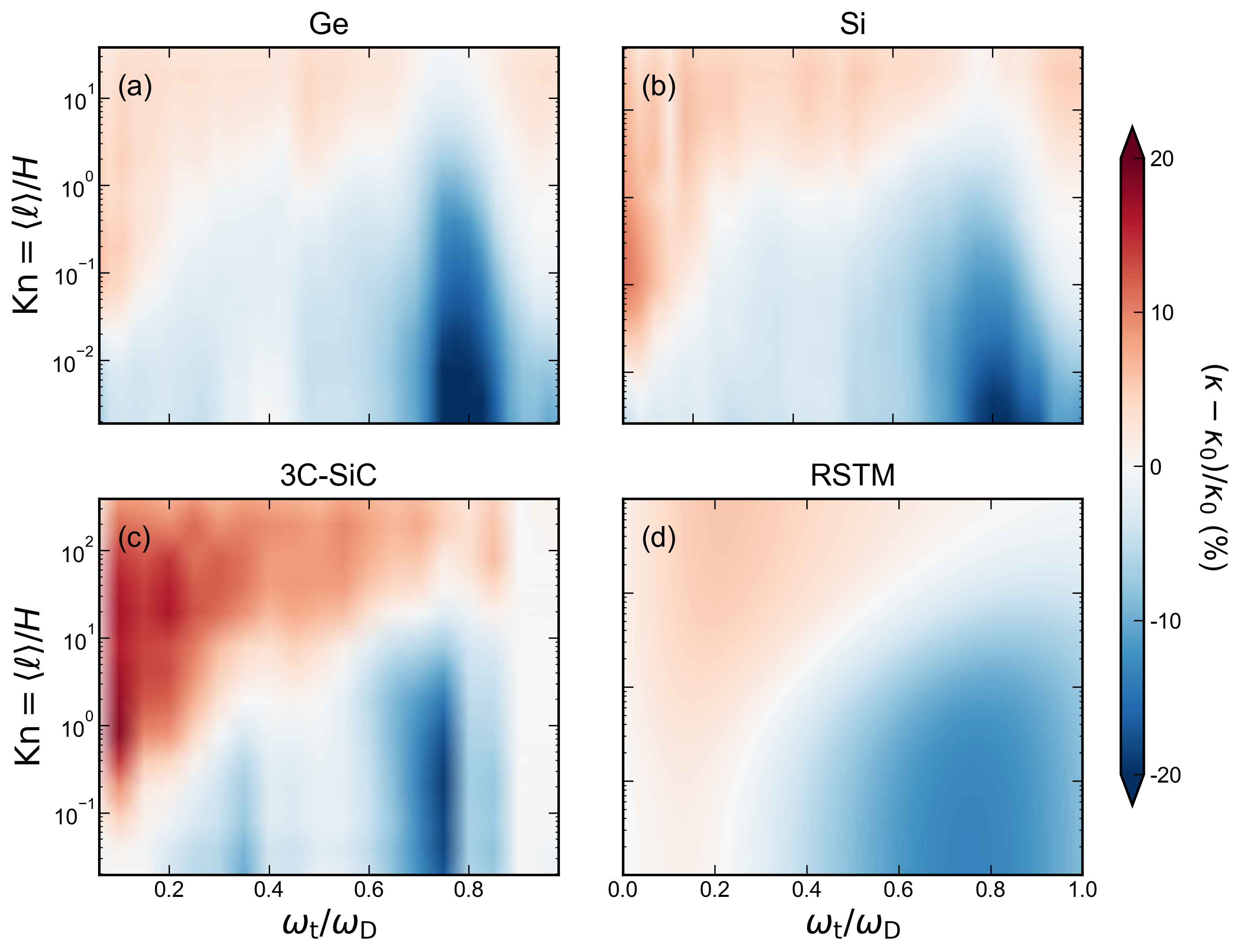}
    \caption{
    \textbf{First-principles-based Monte Carlo and reduced spectral transport model (RSTM) maps of the thermal-conductivity response under targeted phonon excitation.}
    The relative thermal-conductivity response $(\kappa-\kappa_0)/\kappa_0$ is shown as a function of the normalized target frequency $\omega_{\mathrm t}/\omega_{\mathrm D}$ and the Knudsen number $\mathrm{Kn}=\langle\ell^0\rangle/H$.
    Panels (a)--(c) correspond to a background temperature of $T=300~\mathrm{K}$ and an injected energy density of $\Delta E_3=2.0\times10^7~\mathrm{J\,m^{-3}}$.
    Panel (d) is calculated from Eq.~\eqref{eq:total_spectral_response} using the dimensionless spectral functions and parameters of the RSTM specified in SI Appendix, Sec.~S5.
    (a)--(c) Phase maps obtained from first-principles three-phonon scattering rates combined with phonon-tracking Monte Carlo simulations for Ge, Si, and 3C--SiC, respectively.
    The material-specific cutoff frequencies, expressed as $\omega_{\mathrm D}/2\pi$, are 7.79 THz for Ge, 13.34 THz for Si, and 25.00 THz for 3C--SiC.
    (d) Response calculated using the RSTM. 
    Red and blue regions denote excitation-induced enhancement and suppression, respectively, whereas near-white regions indicate the crossover between the two responses.}
    \label{fig4}
\end{figure*}

At a fixed background temperature of $T=300~\mathrm{K}$ and an injected energy density of $\Delta E_3=2.0\times10^7~\mathrm{J\,m^{-3}}$, Figs.~\ref{fig4}(a)--\ref{fig4}(c) organize the material-specific frequency- and confinement-dependent responses into two-dimensional maps. Figure~\ref{fig4}(d) shows the corresponding response map calculated from Eq.~\eqref{eq:total_spectral_response} using the RSTM.

Despite their different phonon spectra and intrinsic scattering strengths, the three materials exhibit a common phase-map topology. Enhancement is concentrated mainly at low normalized target frequencies and within a finite-Knudsen window, where the increased transport weight of long-mean-free-path phonons can outweigh excitation-enhanced intrinsic scattering. Suppression dominates at higher normalized frequencies and in the low-Knudsen bulk-like regime, where intrinsic scattering directly controls the phonon propagation lengths.

For Ge and Si [Figs.~\ref{fig4}(a) and \ref{fig4}(b)], the enhancement region is relatively weak and restricted primarily to low values of $\omega_{\mathrm t}/\omega_{\mathrm D}$. As the target frequency increases, the response changes from positive or nearly neutral to negative, with the strongest suppression occurring near $\omega_{\mathrm t}/\omega_{\mathrm D}\approx0.75$--$0.85$ in the low-Knudsen regime. At these frequencies, excitation provides little additional transport weight to long-propagating modes while substantially increasing intrinsic phonon--phonon scattering.

The 3C--SiC map [Fig.~\ref{fig4}(c)] exhibits a broader and stronger enhancement region. For $\omega_{\mathrm t}/\omega_{\mathrm D}\lesssim0.3$, the response remains positive over a wide range of Knudsen numbers and reaches its largest values within the finite-Knudsen regime. At higher normalized frequencies, the response also becomes negative, showing that sufficiently high-frequency excitation remains predominantly scattering dominated even in the material with the strongest low-frequency enhancement.

The common topology of the three first-principles-based Monte Carlo maps indicates that the response reversal arises from a shared physical competition rather than from a feature unique to one material. To isolate this competition, a RSTM is formulated that separates the excitation-induced increase in modal transport weight from the accompanying increase in intrinsic phonon--phonon scattering. The model is not fitted separately to Ge, Si, or 3C--SiC, and is used to test whether these two ingredients, together with geometric confinement, are sufficient to reproduce the observed response topology.

\subsection*{Reduced spectral transport model (RSTM)}

The RSTM is based on three simplifying assumptions. First, the full wave-vector- and branch-resolved phonon properties are reduced into frequency-dependent spectral functions. Second, the excitation-induced modifications are parameterized by two frequency-dependent enhancement factors. Third, intrinsic and boundary scattering are treated as independent relaxation channels combined through Matthiessen's rule. The RSTM is therefore intended to identify the sign and general topology of the response rather than reproduce the material-specific response magnitude.

Within this reduced description, the competition between excitation-induced enhancement of the modal transport weight and intrinsic scattering is represented in terms of the physical angular frequency $\omega$. For comparison among materials with different phonon frequency ranges, the target frequency is reported in the normalized form $\omega_{\mathrm t}/\omega_{\mathrm D}$, where $\omega_{\mathrm D}$ is the characteristic cutoff angular frequency used to normalize the model spectrum.

Starting from the kinetic spectral representation of lattice thermal conductivity obtained from the phonon Boltzmann transport equation under a relaxation-time description, as discussed on p.~245 of Ref.~\cite{chen_nanoscale_2005} and p.~192 of Ref.~\cite{kaviany_heat_transfer_2014}, let $\mathcal{K}^0(\omega)$ and $\mathcal{K}(\omega,\omega_{\mathrm t})$ denote the equilibrium and excitation-modified spectral conductivity densities, respectively. Up to a common normalization factor, the equilibrium spectral conductivity density is written as
\begin{equation}
\mathcal{K}^0(\omega)
\propto
D(\omega)
C^0(\omega)
v_x^2(\omega)
\tau_{\mathrm{eff}}^0(\omega),
\label{eq:spectral_conductivity_equilibrium}
\end{equation}
whereas the excitation-modified spectral conductivity density is
\begin{equation}
\mathcal{K}(\omega,\omega_{\mathrm t})
\propto
D(\omega)
C(\omega,\omega_{\mathrm t})
v_x^2(\omega)
\tau_{\mathrm{eff}}(\omega,\omega_{\mathrm t}).
\label{eq:spectral_conductivity_excited}
\end{equation}
Here, $D(\omega)$ is the phonon density of states, $v_x(\omega)$ is the group-velocity component along the transport direction, and $C^0(\omega)=\hbar\omega\,\partial f^0/\partial T$ is the equilibrium modal heat capacity. Throughout this subsection, the superscript $0$ denotes the equilibrium reference state, whereas quantities without a superscript refer to the excited state.

Targeted excitation increases the modal transport weight near the target frequency while simultaneously enhancing intrinsic phonon--phonon scattering. The excitation-modified transport weight is represented by
\begin{equation}
C(\omega,\omega_{\mathrm t})
=
C^0(\omega)
\left[
1+\alpha(\omega,\omega_{\mathrm t})
\right],
\label{eq:excitation_transport_weight}
\end{equation}
where $\alpha(\omega,\omega_{\mathrm t})$ is the relative increase in the modal transport weight. Here, $C(\omega,\omega_{\mathrm t})$ is interpreted as a nonequilibrium transport weight rather than as an equilibrium heat capacity evaluated at an elevated temperature.

The accompanying enhancement of intrinsic phonon--phonon scattering is represented by
\begin{equation}
\tau(\omega,\omega_{\mathrm t})
=
\frac{
\tau^0(\omega)
}{
1+\beta(\omega,\omega_{\mathrm t})
},
\label{eq:excitation_intrinsic_lifetime}
\end{equation}
where $\beta(\omega,\omega_{\mathrm t})$ is the relative increase in the intrinsic scattering rate.

In a film of thickness $H$, intrinsic and boundary scattering are combined through
\begin{equation}
\frac{1}{\tau_{\mathrm{eff}}(\omega,\omega_{\mathrm t})}
=
\frac{1}{\tau(\omega,\omega_{\mathrm t})}
+
\frac{|v_x(\omega)|}{H},
\label{eq:effective_tau_model}
\end{equation}
with the equilibrium expression obtained by replacing $\tau$ with $\tau^0$. Defining the equilibrium intrinsic mean free path as $\ell^0(\omega)=|v_x(\omega)|\tau^0(\omega)$ and using $\mathrm{Kn}=\langle\ell^0\rangle/H$, the frequency-resolved relative response becomes
\begin{equation}
\begin{aligned}
R(\omega;\mathrm{Kn},\omega_{\mathrm t})
&\equiv
\frac{
\mathcal{K}(\omega,\omega_{\mathrm t})
-
\mathcal{K}^0(\omega)
}{
\mathcal{K}^0(\omega)
}\\
&=
\left[
1+\alpha(\omega,\omega_{\mathrm t})
\right]
\frac{
1+
\mathrm{Kn}
\dfrac{\ell^0(\omega)}{\langle\ell^0\rangle}
}{
1+
\beta(\omega,\omega_{\mathrm t})
+
\mathrm{Kn}
\dfrac{\ell^0(\omega)}{\langle\ell^0\rangle}
}
-1.
\end{aligned}
\label{eq:R_spectral}
\end{equation}

Equation~\eqref{eq:R_spectral} shows how confinement changes the competition within each spectral channel. In the bulk-like limit, the response is governed mainly by the balance between the increased transport weight and the reduction in intrinsic lifetime. At finite Knudsen numbers, boundary limitation reduces the sensitivity of the effective propagation length to additional intrinsic scattering, particularly for phonons whose intrinsic mean free paths are long compared with the spectrum-averaged value.

The total response is obtained by integrating the channel-resolved contributions over the phonon spectrum,
\begin{equation}
\frac{
\kappa(\mathrm{Kn},\omega_{\mathrm t})
-
\kappa_0(\mathrm{Kn})
}{
\kappa_0(\mathrm{Kn})
}
=
\frac{
\displaystyle
\int_0^{\omega_{\mathrm D}}
\mathcal{K}^0(\omega;\mathrm{Kn})
R(\omega;\mathrm{Kn},\omega_{\mathrm t})
\,d\omega
}{
\displaystyle
\int_0^{\omega_{\mathrm D}}
\mathcal{K}^0(\omega;\mathrm{Kn})
\,d\omega
}.
\label{eq:total_spectral_response}
\end{equation}
Equation~\eqref{eq:total_spectral_response} defines the total relative
response of the RSTM in the same form as the quantity evaluated in
the first-principles-based Monte Carlo simulations. Its sign is determined by
the equilibrium-conductivity-weighted balance of positive and negative
spectral contributions rather than by the behavior of an individual frequency channel.
Evaluating Eq.~\eqref{eq:total_spectral_response} over the $\left(\omega_{\mathrm t}/\omega_{\mathrm D},\mathrm{Kn}\right)$ plane using the dimensionless spectral functions and parameters specified in SI Appendix, Sec.~S5 yields the response map shown in Fig.~\ref{fig4}(d).

Using generic spectral functions, the model reproduces the principal topology of the material-specific maps: enhancement occurs mainly at low normalized target frequencies within a finite-Knudsen window, whereas suppression dominates at higher target frequencies and in the bulk-like regime. Because the model is constructed independently of the detailed phonon properties of Ge, Si, and 3C--SiC, this agreement supports the generic character of the confinement-controlled competition. The first-principles-based Monte Carlo calculations determine the response magnitude and material-specific crossover location, whereas the RSTM identifies the physical mechanism responsible for the common sign reversal.

The explicit dimensionless spectral functions, parameter values, and numerical integration procedure used to generate Fig.~\ref{fig4}(d) are provided in SI Appendix, Sec.~S5.

The microscopic basis of the model is further supported by the mode-resolved relaxation times and cumulative thermal conductivities of the 100-nm Ge, Si, and 3C--SiC films (SI Appendix, Fig.~S4). Low-frequency excitation predominantly affects long-lived acoustic modes that contribute strongly to the cumulative heat current, whereas higher-frequency excitation produces a broader reduction in phonon lifetimes and suppresses the cumulative conductivity.

\section*{Discussion}

The total thermal-conductivity response is governed by the balance over the full phonon spectrum rather than by the response of any individual frequency channel. As expressed by Eq.~\eqref{eq:total_spectral_response}, enhancement occurs only when the equilibrium-conductivity-weighted positive contributions from excitation-modified long-mean-free-path modes exceed the integrated suppression caused by enhanced intrinsic scattering. Geometric confinement shifts this balance by reducing the relative transport penalty associated with a further decrease in intrinsic phonon lifetimes.

This mechanism explains why the largest enhancement occurs at intermediate confinement rather than in either limiting regime. In the bulk limit, the effective propagation lengths are controlled directly by the intrinsic lifetimes, so excitation-enhanced scattering predominantly suppresses thermal conductivity. Under extreme confinement, the propagation lengths of long-mean-free-path phonons are already strongly truncated by the boundaries, limiting the benefit of their increased transport weight. The largest positive response consequently appears when the system size is comparable in scale to the relevant intrinsic mean free paths, where boundary limitation weakens the effect of additional intrinsic scattering without eliminating the contribution of long-propagating modes.

The frequency dependence follows from the spectral properties of the excited phonons. Low-frequency acoustic modes generally possess relatively large group velocities and long intrinsic mean free paths, allowing an increase in their transport weight to produce an appreciable contribution to heat conduction. Higher-frequency modes typically have shorter propagation lengths and contribute less effectively to quasi-ballistic transport. Increasing their population therefore provides comparatively little transport gain while enhancing the occupation-weighted scattering rates experienced across the heat-carrying phonon population. Higher-frequency excitation is consequently predominantly suppressive, although weak positive responses may persist under extreme confinement in some materials.

The common response topology obtained for Ge, Si, and 3C--SiC, together with the independently constructed RSTM, supports a shared confinement-controlled mechanism. The material-dependent phonon spectrum, scattering rates, and mean-free-path distribution determine the magnitude and precise position of the crossover, but the underlying competition remains the same. This active population control is distinct from conventional static size engineering: boundary scattering does not merely reduce thermal conductivity, but changes the susceptibility of heat transport to an externally maintained nonequilibrium phonon population. Film thickness therefore acts both as a passive transport length scale and as a control parameter for the sign of the excitation-induced response.

From an experimental perspective, the normalized target frequency and the Knudsen number represent distinct control variables. The spectral position of the nonequilibrium population may be adjusted through resonant mid-infrared or terahertz phonon excitation, whereas the Knudsen number may be tuned through film thickness, characteristic device dimensions, temperature, or material selection. Experiments have demonstrated mode-selective phonon excitation, terahertz-controlled phonon-mediated relaxation, anharmonic energy transfer between phonon modes, and measurable heat-conduction enhancement by nonequilibrium phonon polaritons~\cite{yoshida_experimental_2013,sekiguchi_enhancing_2021,kozina_terahertz-driven_2019,pan_remarkable_2023}. These developments provide a practical route to test the present prediction: sweeping either the excitation frequency or the Knudsen number across the predicted crossover should reverse the sign of the thermal-conductivity response.

The present framework describes a prescribed quasisteady nonequilibrium phonon population at a fixed background lattice temperature. It therefore addresses the thermal-transport response after the excitation distribution has been established, rather than the time-dependent process through which the population is generated and subsequently relaxes. Coupling to electrons, photons, defects, and other external degrees of freedom is not explicitly included, and sufficiently strong excitation may introduce higher-order scattering, temperature redistribution, or structural changes. Future extensions to time-dependent and experiment-specific excitation mechanisms will be required to predict switching times, relaxation dynamics, excitation powers, and achievable modulation amplitudes. Nevertheless, the present results show that geometric confinement can reverse the thermal-conductivity response to spectrally selective phonon excitation by changing the full-spectrum balance between increased transport weight and enhanced intrinsic scattering.

\section*{Methods}

Targeted phonon excitation is represented by a prescribed quasisteady nonequilibrium population excess maintained within a narrow frequency window centered at the target angular frequency $\omega_{\mathrm t}$. For each phonon mode $\lambda=(\mathbf q,s)$, where $\mathbf q$ and $s$ denote the wave vector and branch index, respectively, the excitation-modified population is written as
\begin{equation}
f_\lambda^{\mathrm{exc}}
=
f_\lambda^0
\left(
1+\alpha_\lambda
\right),
\label{eq:excited_population}
\end{equation}
where $f_\lambda^0$ is the equilibrium Bose--Einstein population at the background lattice temperature. The mode-dependent enhancement factor is prescribed as
\begin{equation}
\alpha_\lambda
=
A
\exp
\left[
-\frac{
\left(
\omega_\lambda-\omega_{\mathrm t}
\right)^2
}{
2\sigma^2
}
\right],
\label{eq:alpha_lambda}
\end{equation}
where $A$ is the excitation amplitude and $\sigma$ is the spectral width in angular-frequency units.

The spectral width is specified by the full width at half maximum,
\begin{equation}
\Delta\left(\frac{\omega}{2\pi}\right)_{\mathrm{FWHM}}
=
\frac{\sqrt{8\ln2}\,\sigma}{2\pi}
=
0.2~\mathrm{THz}.
\label{eq:fwhm_methods}
\end{equation}
All target frequencies reported in THz are therefore expressed directly as $\omega_{\mathrm t}/2\pi$, whereas the normalized frequency used in the phase diagrams is $\omega_{\mathrm t}/\omega_{\mathrm D}$.

The excitation amplitude $A$ is determined by prescribing the injected energy density $\Delta E$. The excess energy density associated with the imposed population is
\begin{equation}
\Delta E
=
\frac{1}{
N_qV_{\mathrm{u.c.}}
}
\sum_\lambda
\hbar\omega_\lambda
f_\lambda^0
\alpha_\lambda,
\label{eq:delta_E_methods}
\end{equation}
where $N_q$ is the number of sampled wave vectors and $V_{\mathrm{u.c.}}$ is the unit-cell volume. Substitution of Eq.~\eqref{eq:alpha_lambda} gives
\begin{equation}
A
=
\frac{
\Delta E\,N_qV_{\mathrm{u.c.}}
}{
\displaystyle
\sum_\lambda
\hbar\omega_\lambda
f_\lambda^0
\exp
\left[
-\frac{
\left(
\omega_\lambda-\omega_{\mathrm t}
\right)^2
}{
2\sigma^2
}
\right]
}.
\label{eq:A_methods}
\end{equation}

The unexcited reference state corresponds to $\Delta E=0$, whereas the excitation-modified calculations use $\Delta E_1=5.0\times10^6~\mathrm{J\,m^{-3}}$, $\Delta E_2=1.0\times10^7~\mathrm{J\,m^{-3}}$, and $\Delta E_3=2.0\times10^7~\mathrm{J\,m^{-3}}$. Prescribing $\Delta E$ rather than a common value of $A$ ensures that different target frequencies are compared at the same injected energy density.

The imposed population is treated as a quasisteady nonequilibrium state maintained by an external energy source, rather than as a Bose--Einstein distribution at an elevated effective temperature. For each target frequency, injected energy density, and background temperature, the excitation profile is held fixed while the thermal-transport response is evaluated. The calculated $\kappa$ therefore represents an effective thermal conductivity around the maintained driven state and does not describe the free temporal relaxation of an isolated nonequilibrium phonon population.

The harmonic and anharmonic phonon properties of bulk Ge, Si, and 3C--SiC are obtained from density-functional-theory calculations performed with Quantum ESPRESSO~\cite{giannozzi_advanced_2017}. The second- and third-order interatomic force constants are used to calculate the phonon frequencies, group velocities, modal heat capacities, and intrinsic three-phonon scattering rates with ShengBTE~\cite{li_shengbte_2014}.

The equilibrium phonon frequencies, group velocities, anharmonic interaction matrix elements, and energy--momentum conservation conditions are used for both the reference and excitation-modified states. Within the calculation of the three-phonon scattering rates, targeted excitation modifies only the Bose occupation factors entering the absorption and emission probabilities. Replacing $f_\lambda^0$ by $f_\lambda^{\mathrm{exc}}$ yields the excitation-modified scattering rate $\Gamma_\lambda^{\mathrm{exc}}$ and intrinsic lifetime
\begin{equation}
\tau_\lambda^{\mathrm{exc}}
=
\frac{1}{
\Gamma_\lambda^{\mathrm{exc}}
}.
\label{eq:tau_exc_methods}
\end{equation}
The corresponding relative enhancement of the intrinsic scattering rate is
\begin{equation}
\beta_\lambda
=
\frac{
\Gamma_\lambda^{\mathrm{exc}}
}{
\Gamma_\lambda^0
}
-1
=
\frac{
\tau_\lambda^0
}{
\tau_\lambda^{\mathrm{exc}}
}
-1.
\label{eq:beta_methods}
\end{equation}

Because the harmonic phonon properties remain unchanged, the difference between the reference and excitation-modified calculations arises from two effects: the increased modal transport weight and the excitation-modified intrinsic lifetime. Under the weak probing temperature gradient, the imposed excitation profile
$\alpha_\lambda$ is held fixed. Within this frozen-profile approximation, the
excitation-modified modal transport weight is defined as
\begin{equation}
C_\lambda^{\mathrm{exc}}
=
C_\lambda^0
\left(
1+\alpha_\lambda
\right),
\label{eq:C_exc_methods}
\end{equation}
which is the mode-resolved counterpart of Eq.~\eqref{eq:excitation_transport_weight}. Here, $C_\lambda^{\mathrm{exc}}$ is a nonequilibrium transport weight rather than an equilibrium heat capacity evaluated at a different temperature.

Phonon modes are sampled according to their normalized modal transport weights,
\begin{equation}
P_\lambda^0
=
\frac{
C_\lambda^0
}{
\displaystyle
\sum_{\lambda'}C_{\lambda'}^0
},
\qquad
P_\lambda^{\mathrm{exc}}
=
\frac{
C_\lambda^{\mathrm{exc}}
}{
\displaystyle
\sum_{\lambda'}C_{\lambda'}^{\mathrm{exc}}
}.
\label{eq:sampling_probabilities}
\end{equation}
The sampled modes are propagated using a phonon-tracking Monte Carlo procedure~\cite{peterson_direct_1994,mazumder_monte_2001,anufriev_ray_2020,liu_quantifying_2025,zhou_effect_2026}. Each trajectory is initialized with a sampled mode $\lambda$ and propagated using its mode-resolved group velocity $\mathbf v_\lambda$. Intrinsic free-flight times are sampled using $\tau_\lambda^0$ for the reference state and $\tau_\lambda^{\mathrm{exc}}$ for the excitation-modified state.

In the bulk calculations, phonon propagation is limited by intrinsic phonon--phonon scattering. In the nanofilm calculations, trajectories are additionally interrupted by interactions with the film boundaries. The same geometry and boundary treatment are used for the reference and excitation-modified calculations, so that their difference arises only from the modified phonon population and intrinsic scattering rates.

The Monte Carlo trajectories are used to determine the mode-resolved effective transport mean free paths $\ell_{\lambda,x}^{\mathrm{eff},0}$ and $\ell_{\lambda,x}^{\mathrm{eff},\mathrm{exc}}$ along the heat-flow direction $x$. These quantities include intrinsic phonon--phonon scattering and, in nanofilms, geometric boundary limitation. They are therefore distinguished from the intrinsic mean free paths $\ell_\lambda^0=|v_{\lambda,x}|\tau_\lambda^0$ used to define the Knudsen number and construct the RSTM.

The total thermal conductivities are reconstructed from the mode-resolved effective transport lengths as
\begin{equation}
\kappa_0
=
\frac{1}{
N_qV_{\mathrm{u.c.}}
}
\sum_\lambda
C_\lambda^0
|v_{\lambda,x}|
\ell_{\lambda,x}^{\mathrm{eff},0},
\label{eq:kappa0_MC}
\end{equation}
and
\begin{equation}
\kappa
=
\frac{1}{
N_qV_{\mathrm{u.c.}}
}
\sum_\lambda
C_\lambda^{\mathrm{exc}}
|v_{\lambda,x}|
\ell_{\lambda,x}^{\mathrm{eff},\mathrm{exc}}.
\label{eq:kappa_exc_MC}
\end{equation}
The excitation-modified transport weight and effective mean free path play distinct roles in these expressions. The former accounts for the increased contribution of the selectively populated modes, whereas the latter accounts for excitation-enhanced intrinsic scattering and geometric boundary limitation. The phonon-population modification is therefore not counted twice. The relative response is calculated from $\kappa_0$ and $\kappa$ using Eq.~\eqref{eq:relative_response}.

For each material, the response is evaluated over the sampled values of $\omega_{\mathrm t}/2\pi$, film thickness, background temperature, and injected energy density. The results are subsequently mapped onto the $\left(\omega_{\mathrm t}/\omega_{\mathrm D},\mathrm{Kn}\right)$ plane.
The same equilibrium definition of $\langle\ell^0\rangle$ is used for all film thicknesses at a given material and temperature.

The material-specific cutoff frequencies used for normalization are $\omega_{\mathrm D}/2\pi=7.79~\mathrm{THz}$ for Ge, $13.34~\mathrm{THz}$ for Si, and $25.00~\mathrm{THz}$ for 3C--SiC. Here, $\omega_{\mathrm D}$ denotes the material-specific upper cutoff of the calculated phonon spectrum. The first-principles-based Monte Carlo maps in Figs.~\ref{fig4}(a)--\ref{fig4}(c) are obtained by interpolating the calculated values of $(\kappa-\kappa_0)/\kappa_0$ in the $\left(\omega_{\mathrm t}/\omega_{\mathrm D},\mathrm{Kn}\right)$ plane.

The RSTM used to generate Fig.~\ref{fig4}(d) is conceptually and computationally distinct from the first-principles-based Monte Carlo calculations. The material-specific maps in Figs.~\ref{fig4}(a)--\ref{fig4}(c) retain the calculated phonon dispersions, mode-dependent three-phonon scattering rates, and trajectory-level boundary interactions of Ge, Si, and 3C--SiC. By contrast, the RSTM uses generic spectral functions for $D(\omega)$, $v_x(\omega)$, $\ell^0(\omega)$, $\alpha(\omega,\omega_{\mathrm t})$, and $\beta(\omega,\omega_{\mathrm t})$ to isolate the competition between increased modal transport weight and enhanced intrinsic scattering. The model is not fitted separately to the three materials and is used only to determine whether these ingredients are sufficient to reproduce the common response-map topology.

The RSTM map in Fig.~\ref{fig4}(d) is calculated from Eq.~\eqref{eq:total_spectral_response}. The explicit dimensionless spectral functions, parameter values, and numerical integration procedure used to generate Fig.~\ref{fig4}(d) are provided in
SI Appendix, Sec.~S5.

\begin{acknowledgments}

This work was supported by Innovation Research Foundation of National University of Defense Technology (Innovation Research Foundation of NUDT).
S. Liu and F. Yin acknowledge the China Scholarship Council (No. 202308090243 for S. Liu and No. 202408090635 for F. Yin).
\end{acknowledgments}

\bibliography{myref}

\end{document}